\documentclass{article}
\usepackage{spconf,amsmath,epsfig}

\usepackage[numbers]{natbib}
\usepackage[caption=false,font=footnotesize]{subfig}

\title{The statistics of radio frequency interference propagating from long distances to the Murchison Radio-astronomy Observatory}
\twoauthors
  {Marcin Sokolowski\sthanks{marcin.sokolowski@curtin.edu.au}, Randall B. Wayth}
 	{ICRAR/Curtin University\\
 	GPO Box U1987\\Perth, WA, 6845. Australia\\
 	ARC Centre of Excellence for\\
        All-Sky Astrophysics (CAASTRO)}
  {Tyler Ellement}
 	{Curtin University\\
 	 GPO Box U1987\\Perth, WA, 6845. Australia}
\begin{document}
%
\maketitle
\begin{abstract}
BIGHORNS is a total power radiometer developed to identify the signal from the Epoch of Reionisation in the all-sky averaged radio spectrum at low frequencies (70-300 MHz). 
In October 2014, the system with a conical log spiral antenna was deployed at the Murchison Radio-astronomy Observatory (MRO) and has been collecting data since then. 
The system has been monitoring the radio frequency interference (RFI) environment at the future site of the low-frequency component of the Square Kilometre Array. 
We have analyzed almost two years of data (October 2014 - August 2016 inclusive) in search for events of long distance propagation of the RFI in FM and digital TV bands due to tropospheric ducting and reflections (of the meteor trails or aircraft). 
We present statistics of tropospheric ducting events observed in the digital TV band over nearly two years, which shows seasonal changes. 
We also present a system using upper atmosphere data (temperature, humidity and pressure as a function of altitude) from all stations in Western Australia to calculate the modified refractive index and make predictions of tropospheric ducting events. 
Preliminary tests indicate that the system can be very useful in predicting tropospheric ducting events (even with limited amount of available upper atmosphere data).
\end{abstract}
\section{Introduction}
\label{sec:intro}

Radio-frequency interference (RFI) is an ongoing challenge for radio astronomy because the difference in received power between astronomical radio sources and human-generated sources can be many orders of magnitude.
The effects of RFI range from the extreme case where the receiver gets saturated rendering all data unusable, through moderate case in which the affected frequency channels can be flagged and excluded from further analysis down to low power RFI which might be very hard to detect and only manifests itself in high sensitivity measurements (requiring very long integration times).
Paradoxically, strong RFI can be easier to handle as it is easy to identify, flag and can be excluded from further analysis whilst low power RFI can potentially remain in the data significantly distorting astronomical observations.
Multiple science programs of current and future radio telescopes, including the Square Kilometre Array (SKA), require thousands of hours of observing time \citep[e.g][]{2013PASA...30...31B,2015aska.confE...1K}, which can be significantly affected by even very low power RFI.

Therefore, the low frequency component of the SKA (SKA-low) will be built in the Murchison Radio-astronomy Observatory (MRO), which is located in a remote area of Western Australia (WA) where a Australian Radio Quiet Zone - Western Australia (ARQZWA) \citep{rqz} has been established in order to ensure high quality data. 
There are currently several low-frequency radio instruments operating at the MRO including the Murchison Widefield Array (MWA) \cite{2013PASA...30....7T} a precursor instrument for the future SKA-Low telescope \cite{2015arXiv151001515S}, several other test installations for the SKA-low telescope (AAVS-1 and EDA) and BIGHORNS \cite{2015PASA...32....4S}.
Although the MRO is in a designated radio quiet zone (RQZ) with significant regulatory protections, RFI from aircraft and satellite-based transmitters is detectable as well as occasional weak FM radio, digital TV (DTV) or digital radio (DR) broadcasts from distant transmitters.
Particularly, long distance propagation of FM and DTV signals due to tropospheric ducting (TD) \cite{tropo_ducting} or reflections can cause excision of significant fractions of scientifically useful bands. 
Especially, ducted DTV and DR propagated from far away transmitters (even 600\,km from Perth) can potentially cause significant data loss due to its broadband nature and long timescale (usually several hours during the nighttime).

Thus, it is important to study and understand the statistics, seasonal and daily patterns of long distance RFI propagation in order to optimize observation planning and potentially implement additional data quality checks based on meteorology data
allowing to predict TD events.

\begin{figure*}[]
\centering
\includegraphics[width=\textwidth]{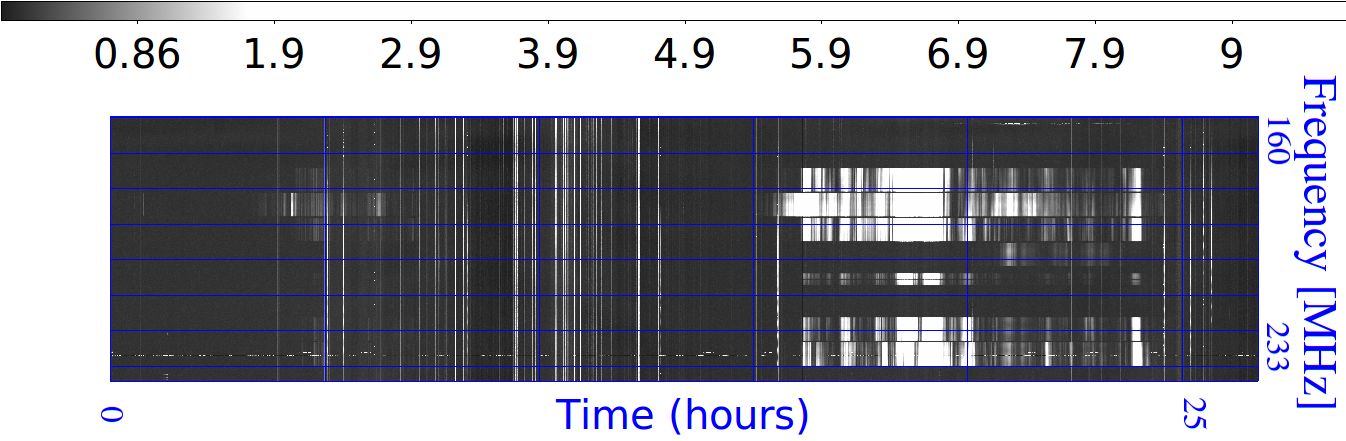}
\caption{Normalized dynamic spectrum in dB scale collected by BIGHORNS between 2014-10-29 00:00 and 2014-10-30 12:00 AWST. The right part of the spectrum (night 2014-10-29/30) shows a strong tropospheric ducting event observed in DTV band with RFI propagated from Perth (some 600\,km from the MRO).}
\label{fig:bighorns_ducting_noducting}
\end{figure*}

This paper is organized as follows. In Section~\ref{sec:data_daq_and_proc} we give a brief overview of the BIGHORNS total power radiometer, data acquisition and processing.
In Section~\ref{sec:results} we present statistics of the tropospheric ducting events observed with BIGHORNS system between Oct 2014 and August 2016 (inclusive). We classified the identified TD events as strong, moderate and weak based on DTV signals (180-220\,MHz). 
We also present preliminary tests of a system using upper atmosphere meteorology data from stations in Western Australia to predict TD events. 
Finally, in Section~\ref{sec:summary} we summarize and present possible further steps and improvements in our TD prediction system.


\section{Data acquisition and processing}
\label{sec:data_daq_and_proc}

\subsection{BIGHORNS system}
\label{subsec:data_acq}

The data were recorded with the BIGHORNS system. The details of the system can be found in \citep{2015PASA...32....4S,2015ApJ...813...18S} and we only give a very brief summary here.
BIGHORNS is a calibrated total power spectrometer attached to a single antenna. 
For most of the observing period (October 2014 - August 2016) a bespoke conical log-spiral (CLS) antenna well matched to left-hand circularly  polarized (LHCP) 
emission from the sky between 50 and 350\,MHz was used \cite{2015ApJ...813...18S} and one month of data (Sep/Oct 2015) was collected with a biconical antenna \citep{2015PASA...32....4S}.
The system records the power detected from the sky with approximately 117.2\,kHz frequency resolution and 0.05\,s time resolution.
Filters in the BIGHORNS signal chain limit the useful upper frequency to approximately 300\,MHz and ripples in the antenna frequency response limit the lower frequency to approximately 70\,MHz.

\subsection{Data processing}
\label{subsec:data_proc}

In order to reduce the amount of data, the raw data from each month were averaged in 35\,sec time bins and calibrated according to the standard procedure developed for the BIGHORNS instrument \cite{2015PASA...32....4S,2015ApJ...813...18S}. 
Each calendar month was analyzed separately. In order to identify daily variations in the sky signal (for example due to propagation in the ionosphere) and RFI, the calibrated spectra were normalized by a median of all the spectra collected at the same local sidereal time (LST) during an analyzed month. 
Without any variations in the system and the ionosphere such a normalized dynamic spectrum should be equal to one (within the noise) as the radio sky is the same for any given LST (see \citep{2015ApJ...813...18S} for details of the normalization procedure). 
The normalized dynamic spectrum was converted into dB scale. Figure~\ref{fig:bighorns_ducting_noducting} shows 36\,hours of normalized dynamic spectrum in dB scale. Then the normalized dynamic spectra from each month were visually inspected to identify tropospheric ducting events in DTV band and classify them as: strong (at least 10\,dB above median background power), moderate (above 3\,dB) and weak (below 3\,dB).

\section{Results}
\label{sec:results}

\subsection{Seasonal and daily patterns in tropospheric ducting}
\label{subsec:seasonal_daily_patterns}

\begin{figure}[]
\centering
\includegraphics[width=0.5\textwidth]{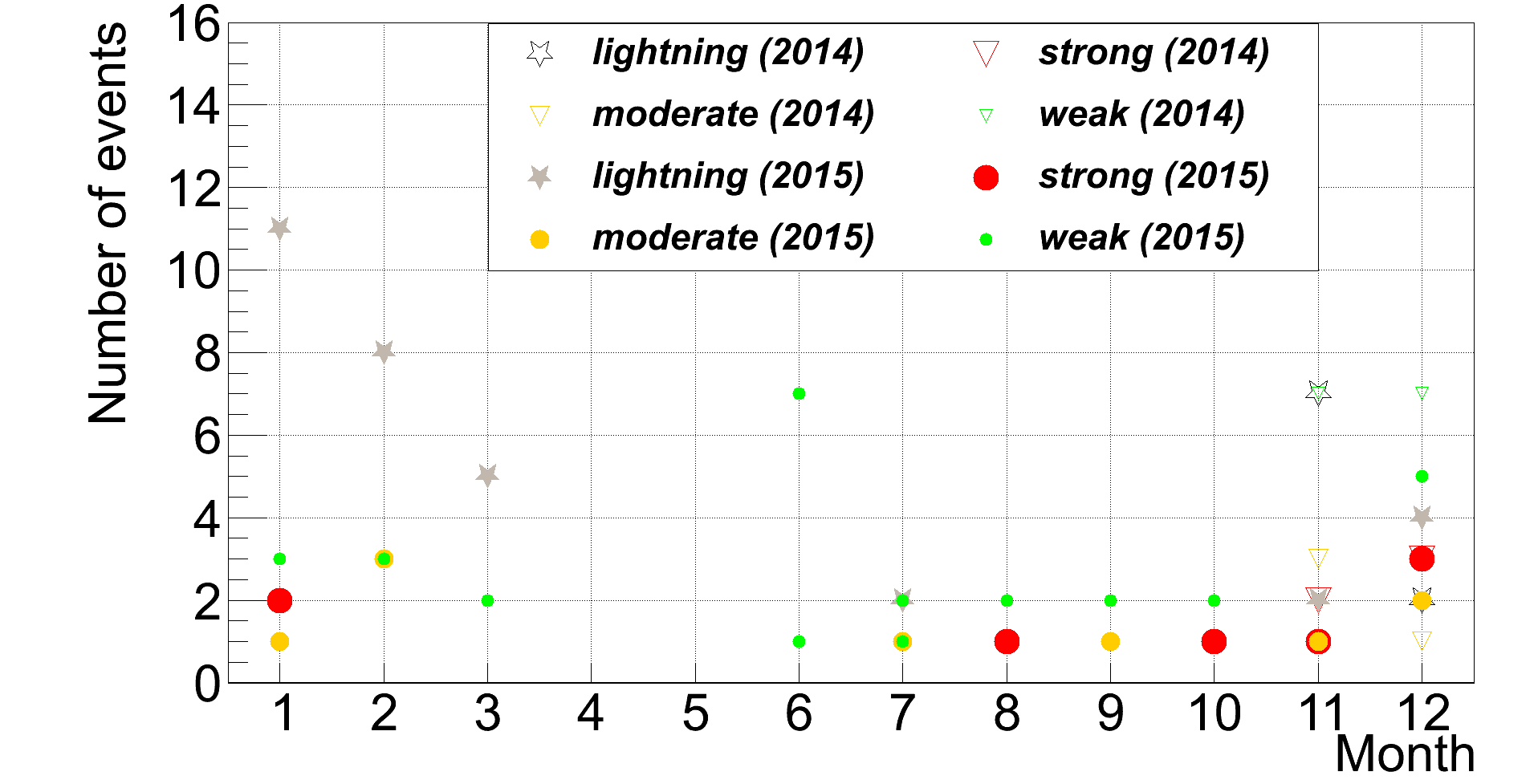}
\caption{Number of tropospheric ducting and thunderstorms at the MRO in 2014 (only November and December) and 2015. During this time no data were collected in the following periods: 2015-03-03 - 2015-03-17, 2015-04-06 - 2015-04-31, 2015-10-09 - 2015-11-24 and several shorter (maximum a few days) gaps due to system downtime.}
\label{fig:ducting_2014_2015}
\end{figure}

\begin{figure}[]
\centering
\includegraphics[width=0.5\textwidth]{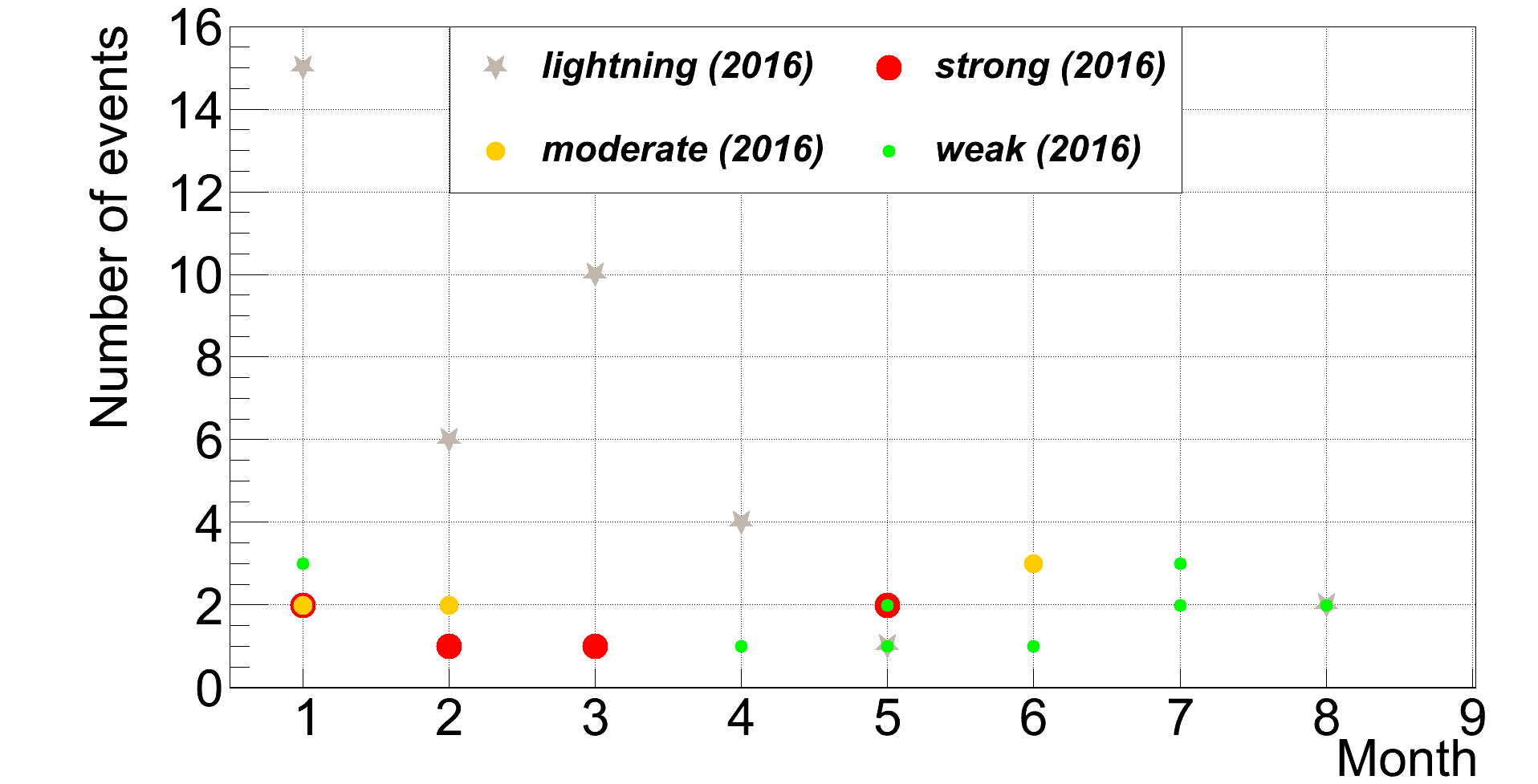}
\caption{Number of tropospheric ducting and thunderstorms at the MRO in 2016 (until 31 August 2016). During this time no data were collected in the following periods: 2016-02-25 - 2016-03-09, 2016-04-05 - 2016-05-04 and 2016-08-17 - 2016-08-29 and several shorter (maximum a few days) gaps due to system downtime.}
\label{fig:ducting_2016}
\end{figure}

We inspected all the data collected between 2014-10-24 and 2016-08-31, which covered 516 days in total because of several gaps due to system downtime (see captions of Figures~\ref{fig:ducting_2014_2015} and \ref{fig:ducting_2016}). 
The TD events were observed in BIGHORNS data only during nighttime. They usually started after sunset or even later during the night and persisted until early morning (not later than 9~am).
During the entire time period we observed 19 strong, 20 moderate and 59 weak tropospheric ducting events (overall almost 20\% of analyzed nights) \footnote{The list of all TD events observed in the analyzed time period is available on request.}.
Typically, good and very good TD conditions persist for 2-3 nights. However, there are also multiple examples (especially during colder months) of isolated events (only one night affected).

We used existing MWA observations from the night 2014-10-29/30 when a strong TD event was observed \citep{bighorns_rfi_paper} to make an all-sky image and identified that the source of the interference was on the horizon in the direction towards Perth (600\,km from the MRO).
The RFI source was present only at the frequencies where strong DTV and DR signals were observed in the corresponding BIGHORNS data.
We are planning to perform more targeted MWA observations of TD events in order to better understand the RFI environment at the MRO and identify possible sources of ducted RFI.

In Figures~\ref{fig:ducting_2014_2015} and \ref{fig:ducting_2016} we present monthly summary of the ducting events for each month in time periods 2014-10-24 - 2015-12-31 and 2016-01-01 - 2016-08-31 respectively.
During the hotter months (October, November, December, January and February) there were typically a few (not more than 3) strong and moderate events and multiple weak TD events. 
It is also noticeable that colder winter months are typically much quieter in this regards with significantly less strong and moderate events. However, there were still several nights when strong tropospheric ducting occurred (for example mornings of 2016-05-11 and 2016-05-24).

Our identification of TD events is limited by the sensitivity of BIGHORNS instrument on 35\,sec integration. Hence, it is possible that extremely weak TD events have been missed.
Therefore, we would like to better understand seasonal and daily patters of tropospheric ducting events and possibly develop an additional data quality criteria (based on upper atmosphere meteorology data). 
Particularly, in the context of high sensitivity measurements requiring very long integrations, such as for example Epoch of Reionisation (EoR) experiment, which might be very susceptible to low level, unflagged RFI.
We have verified that some of the very weak TD events which we identified in BIGHORNS data were not flagged by the RFI flagging software (aoflagger \citep{aoflagger}). 
This is an example of very low level, unflagged RFI ``sneaking'' into the final data sample, which may also happen in the MWA data and thus increase the non-Gaussian noise. However, MWA tile and BIGHORNS CLS antenna beams are different with MWA tile beam much less sensitive at the horizon.
Although near-the-horizon RFI is further suppressed by interferometric ($\sim$30\,dB) formation of the synthesized beam by the MWA, the effect should be considered and potentially additional data quality criterion (using upper atmosphere data) developed to excise data affected by TD.

\subsection{Correlation with higher frequencies}

During the first week of Jan 2016 we observed TD events almost every night with two strong TD during nights 2016-01-04/05 and 2016-01-06/07. Very similar observations were reported in \citep{balt} at higher frequencies (930-960\,MHz) based on data from the Australian Square Kilometre Array Pathfinder (ASKAP) \citep{2007PASA...24..174J}, which is also located at the MRO.
These observations further confirm that the observed events are due to TD which is expected to occur in wide frequency range. We are planning to compare more low and high frequency data in the future.

\subsection{Comparison with tropospheric ducting predictions}
\label{subsec:compare_with_expectations}


\begin{figure*}[!t]
\centering
\subfloat{\includegraphics[width=0.45\textwidth]{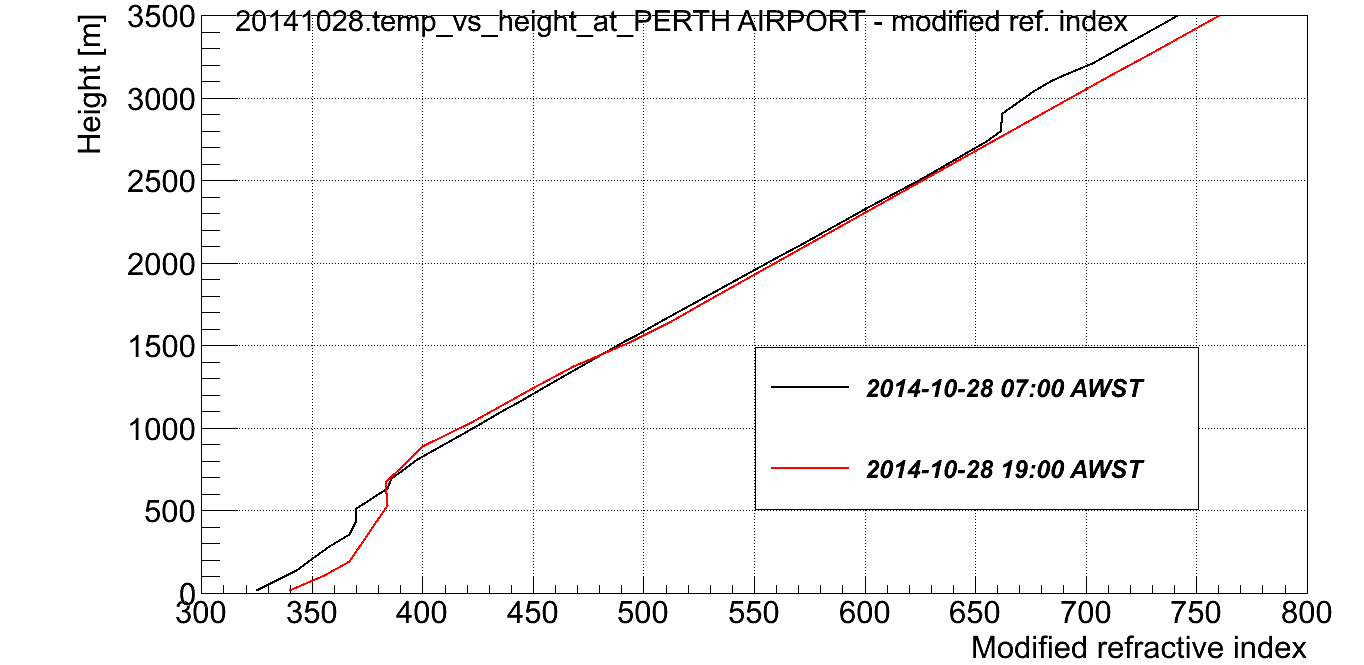}}
\subfloat{\includegraphics[width=0.45\textwidth]{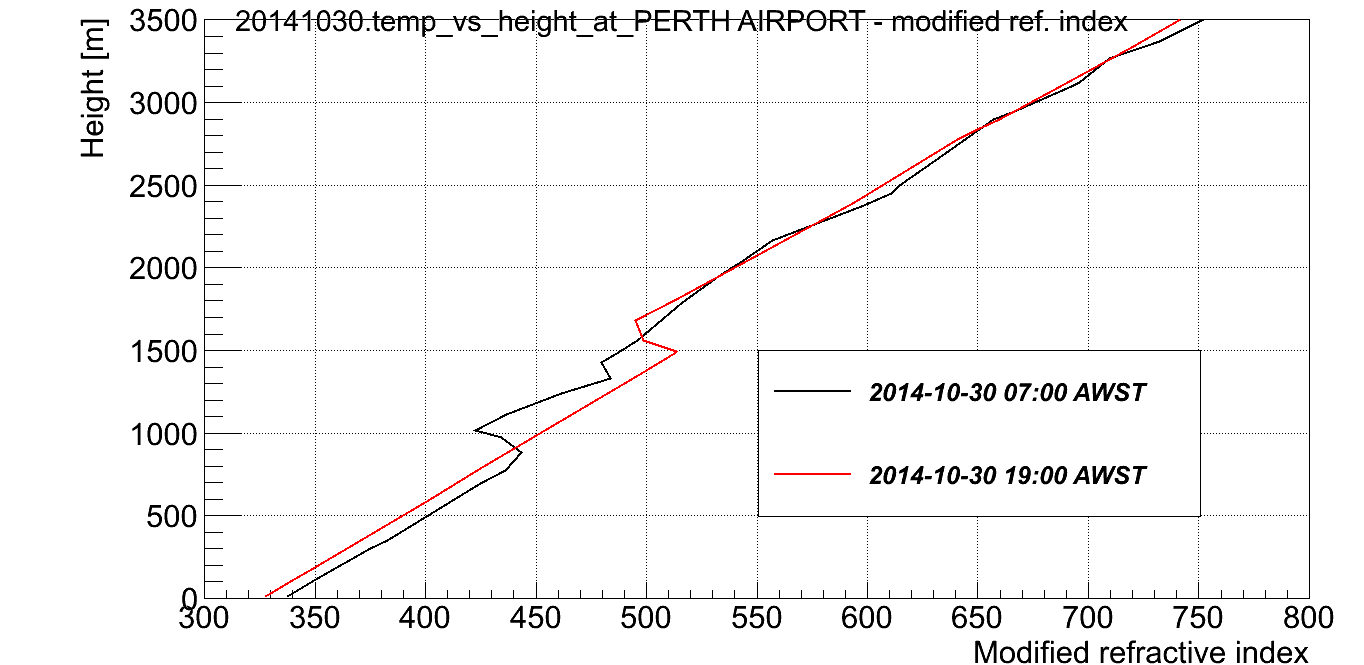}}
\caption{The modified refractive index (MRI) calculated from upper atmosphere data provided by the Perth Airport station on 28th October 2014 when no TD was observed (left image) and 30th October 2014 when strong TD was observed (right image). In the right image we can see a typical signature of good TD conditions with a negative MRI gradient at altitude about 1000\,m (at 7\,am AWST). A similar dependence was observed in MRI calculated from Geraldton and Meekatharra upper atmosphere measurements.}
\label{fig:mir_ducting_and_noducting}
\end{figure*}

A very good review of the tropospheric ducting phenomena can be found in \citep{tropo_ducting}.
Generally, good TD conditions occur when a layer of troposphere has a negative gradient of the modified refractive index (meaning that the radio ray's curvature is higher than Earth's curvature), which is usually related to temperature inversion \citep{tropo_ducting}.
Temperature inversion conditions are more common in the summertime and can be even more common near the coastline.
Depending on the particular values of the gradient of the modified refractive index (MRI) there are several ducting regimes and in extreme case (negative gradient) radio waves can be trapped in a layer of troposphere and form a so called ``tropospheric duct''.
A surface-based duct is formed when the trapping layer is between the ground and the reflecting layer, an elevated duct is formed when the trapping layer is somewhere above the ground and evaporation ducts are formed above oceans and seas due to evaporating water. 

In order to predict TD events MRI needs to be calculated from upper atmosphere (UA) meteorology data (temperature, humidity and pressure as a function of altitude). 
For the analyzed time interval we obtained the UA data from Perth, Geraldton and Meekatharra stations from the Australian Bureau of Meteorology (BOM) \footnote{http://www.bom.gov.au/}
and calculated MRI according to formulas presented in \citep{ITUR_453}, \citep{ITUR_835} and \citep{hajime_suzuki}.
Example of MRI as a function of altitude calculated from UA data from Perth from 2014-10-28 and 2014-10-30 is shown in 
Figure~\ref{fig:mir_ducting_and_noducting} and normalized dynamic spectrum in dB scale from BIGHORNS for the two nights without
(2014-10-28/29) and with (2014-10-29/30) TD is shown in Figure~\ref{fig:bighorns_ducting_noducting}.
The MRI calculated from Geraldton UA data for the same nights is very similar and also indicated good TD conditions. 
These data show that at least in some cases TD can be predicted from the UA data measured at a limited number of stations.

Out of all $\approx$100 TD events observed by BIGHORNS in (October 2014 - August 2016) about 50\% found confirmation in UA data from the three stations (Perth, Geraldton and Meekatharra) collected within 24\,h from the event (about 60\% within 48\,h).
For strong and moderate TD events $\approx$60\% within 24\,h and 75\% within 48\,h were also expected based on the UA data. As can be seen there was always a certain number of TD events observed in BIGHORNS data which could not be explained based on the MRI calculated from UA data from only 3 stations.
This is not unexpected as there are many other places in WA and possibly also other states where ducted RFI could come from (i.e. we think that the ``unexplained'' TD events could come from different locations than Perth, Geraldton and Meekatharra).
We are planning to further study potential sources of ducted RFI by using MWA telescope to image more TD events.

On the other hand, in the same period about 50\% out of about 150 TD events expected from UA data were observed in BIGHORNS data (we excluded downtime periods). 
Some significant loss in TD identification rate in the summertime was due to thunderstorms near the MRO which produced very significant RFI hindering identification of TD events in BIGHORNS data (Figures~\ref{fig:ducting_2014_2015} and \ref{fig:ducting_2016}).

\begin{figure*}[t]
\centering
\includegraphics[width=\textwidth]{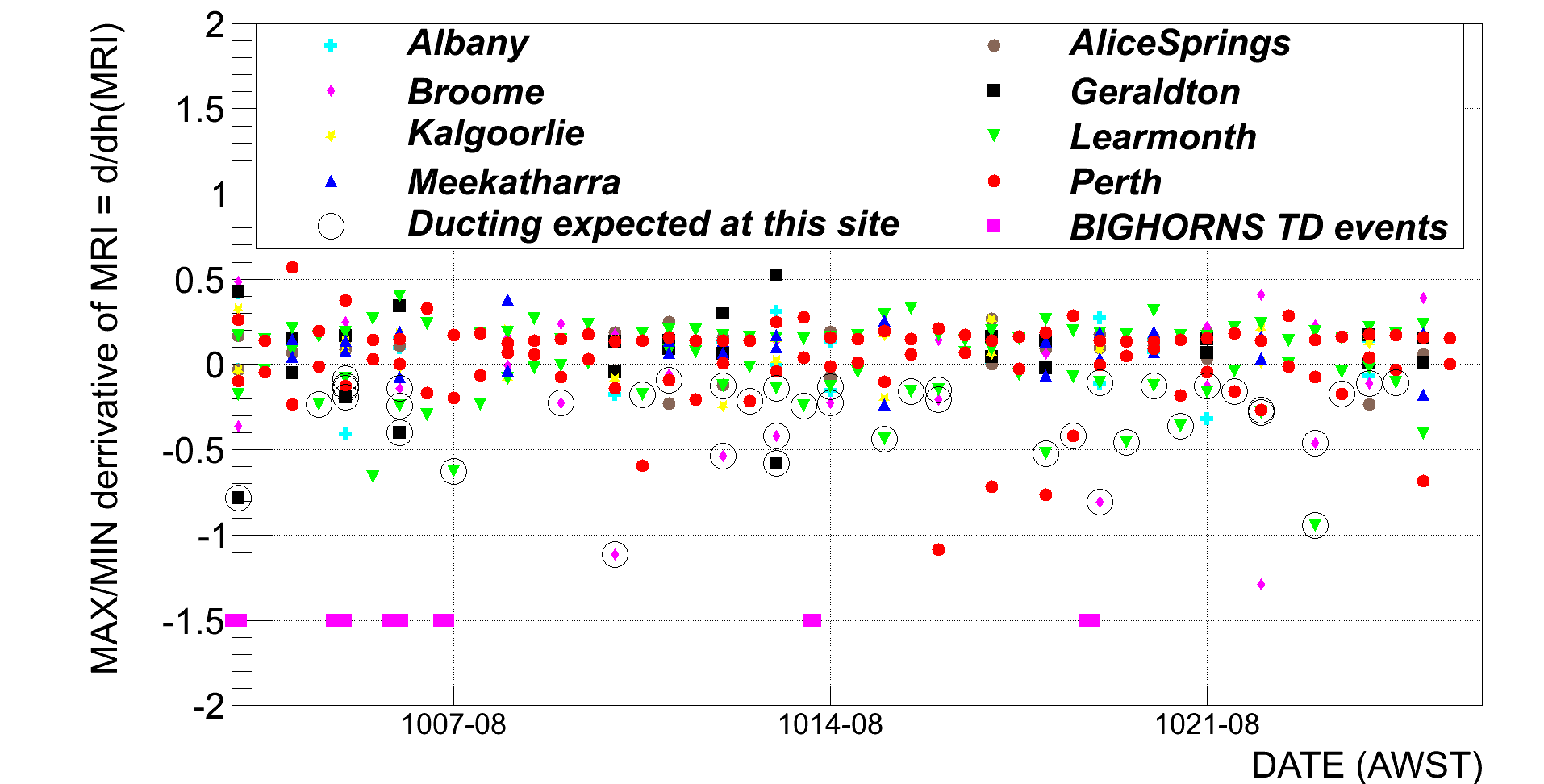}
\caption{Maximum and minimum height derivative of MRI calculated from October 2016 UA data from all stations in WA and Alice Springs. If minimum value of the derivative was below -0.079 M/m at altitude below 1000\,m we considered TD to be expected at a given location. The predicted TD events are marked by an open black circle surrounding the colour data point. Magenta rectangles at arbitrary chosen value -1.5 show TD events observed with BIGHORNS system. All BIGHORNS TD events correlated well with expected TD events calculated from the UA data from Geraldton, except TD event observed on 2016-10-07 at 2-5:30 am AWST when no UA data were collected by the Geraldton station on 7,8 and 9th of October 2016. The TD event observed by BIGHORNS during the night 2016-10-18/19 does not have a corresponding TD prediction from Gerladton station due to limited time coverage of UA measurements in Geraldton (and other remote stations), but the UA data from the Perth station indicates good TD conditions. It can also be noted that TD is much more common at warm coastline locations (Learmonth and Broome).}
\label{fig:predict_vs_data_201610}
\end{figure*}

We have recently implemented daily downloads of the UA data from all WA stations (Perth, Geraldton, Meekatharra, Learmonth, Broome, Kalgoorlie and Albany) and Alice Springs.
So far, for most of the October 2016 nights whenever MRI calculated from Perth or Geraldton data suggested good TD conditions, we indeed observed TD events in the BIGHORNS data and vice-versa \footnote{With the exception of TD event on 2016-10-07 (2-5am) for which there are no UA data from Geraldton} (Fig.~\ref{fig:predict_vs_data_201610}).


\section{Summary}
\label{sec:summary}

We have analyzed almost two years of BIGHORNS archival data (October 2014 - August 2016) and identified 98 tropospheric ducting (TD) events in these data (almost 20\% of the 516 analyzed days). 
Based on the ratio of power observed in digital TV bands to median noise floor at the same frequency, we have classified these events as strong ($>$10\,dB), moderate ($>$3\,dB), weak ($<$3\,dB). We have identified 19 strong, 20 moderate and 59 weak TD events.
We have found that TD events occur only during nighttime (often start just after sunset) and typically last for a couple of hours, sometimes until morning hours and sometimes even after sunrise, but not later than 9-10~am.
Their duration vary from about 1\,hour to many hours. In several cases strong and moderate TD events started just after sunset and continued until morning hours after sunrise.
We identified that the most favorable conditions for TD at MRO are during the hot and warm months (October, November, December, January and February) when usually 1-3 strong and multiple weaker TD events were observed in every month. 
During the warm months very often favorable TD conditions persist for 2 or even 3 consecutive nights. Cold, winter nights are much quieter, but occasionally even strong and moderate isolated TD events were observed. 
Using existing MWA data, we imaged one of the strong TD events (night 2014-10-29/30) to find out that the RFI propagated from Perth (about 600\,km from the MRO).
We are planning to perform more all-sky MWA observations in order to identify the potential sources of RFI due to TD.

We used upper atmosphere (UA) data from Perth, Geraldton and Meekatharra stations obtained from BOM to calculate modified refraction index (MRI) and verify if the observed TD events correlated with the MRI based expectations (when MRI as a function of altitude indicates super-refraction or trapping).
In about 50-60\% cases of the TD events (depending on the class of event) observed with BIGHORNS at the MRO the UA data from the three stations indeed indicated good TD conditions.
Similarly, in approximately 50-60\% cases of UA data indicating good TD conditions we did observe TD in the BIGHORNS data. 
The main reason for this inaccuracy may come from the limited coverage of the UA data (only a couple of locations in WA) and in many cases limited time coverage (for example Geraldton station typically performs measurements every second day).
Unfortunately, it is not possible to obtain measurements from more locations along the line of sight, but we are planning to use interpolation of the existing data to further improve our TD predictions.
We presently automatically download UA data (a few times a day) from all the stations in Western Australia and Alice Springs calculate MRI, make TD predictions and compare with BIGHORNS observations. 
Preliminary, October 2016 results suggest that if MRI altitude structure indicates good TD conditions in Perth or Geraldton (or both) we usually observe TD events with BIGHORNS system at the MRO (Fig.~\ref{fig:predict_vs_data_201610}). 
We would like to further understand the seasonal, daily patterns of TD events and improve our ability to predict these events based on available UA data in order to have a better picture of the observing conditions at the MRO.
Particularly, in the context of developing data quality control for very sensitive measurements (for example EoR), which require many hours of integration and can be significantly affected by even very weak, unflagged RFI.

\section{Acknowledgment}
We would like to thank Hajime Suzuki from CSIRO for his help with using the upper atmosphere data and testing calculation of the modified refractive index.
We thank CSIRO and the MWA operations team for their ongoing support of BIGHORNS and its supporting infrastructure at the Murchison Radio-astronomy Observatory. We acknowledge the Wajarri Yamatji people as the traditional owners of the Observatory site.
This research was conducted by the Australian Research Council Centre of Excellence for All-sky Astrophysics (CAASTRO), through project number CE110001020.
The International Centre for Radio Astronomy Research (ICRAR) is a Joint Venture between Curtin University and the University of Western Australia, funded by the State Government of Western Australia and the Joint Venture partners.


\newcommand{\pasa}{PASA}
\newcommand{\apj}{ApJ}
\newcommand{\aj}{AJ}
\newcommand{\apjl}{ApJ}
\newcommand{\aap}{A\&A}
\newcommand{\mnras}{MNRAS}
\newcommand{\aaps}{A\&AS}
\newcommand{\pasp}{PASP}
\newcommand{\physrep}{PhR}
\newcommand{\araa}{ARA\&A}

\bibliographystyle{IEEEbib} 
\bibliography{refs}

\end{document}